%Paper: gr-qc/9405031
%From: Renate Loll <loll@phys.psu.edu>
%Date: Wed, 11 May 94 20:03:59 -0400

% paper "New loop representations for 2+1 gravity", by A. Ashtekar
% and R. Loll, begins here
%%%%%%%%%%%%%%%%% fonts, definitions, etc. %%%%%%%%%%%%%%%%%%%%%%%%%%%%
\input epsf.tex

\font\rmu=cmr10 scaled\magstephalf
\font\bfu=cmbx10 scaled\magstephalf

\font\it=cmti10 scaled \magstephalf

\rmu

\font\rmus=cmr8
\font\rmuss=cmr6
\font\mait=cmmi10 scaled\magstephalf
\font\maits=cmmi7 scaled\magstephalf
\font\maitss=cmmi7
\font\msyb=cmsy10 scaled\magstephalf
\font\msybs=cmsy8 scaled\magstephalf
\font\msybss=cmsy7
\font\bfus=cmbx7 scaled\magstephalf
\font\bfuss=cmbx7
\font\cmeq=cmex10 scaled\magstephalf

\textfont0=\rmu
\scriptfont0=\rmus
\scriptscriptfont0=\rmuss

\textfont1=\mait
\scriptfont1=\maits
\scriptscriptfont1=\maitss

\textfont2=\msyb
\scriptfont2=\msybs
\scriptscriptfont2=\msybss

\textfont3=\cmeq
\scriptfont3=\cmeq
\scriptscriptfont3=\cmeq

\newfam\bmufam  \textfont\bmufam=\bfu
      \scriptfont\bmufam=\bfus \scriptscriptfont\bmufam=\bfuss

\hsize=15.5cm
\vsize=22cm
\baselineskip=16pt   % Double spacing
\parskip=12pt plus  2pt minus 2pt

\def\a{\alpha}
\def\b{\beta}
\def\d{\delta}
\def\e{\epsilon}

\def\g{\gamma}

\def\A{\cal A}
\def\G{\cal G}
\def\ag{\A/\G}
\def\agb{\overline{\A/\G}}
\def\semi{\bigcirc\kern-1em{s}\;}

\def\del{\partial}
\def\ni{\noindent}
\def\R{{\rm I\!R}}

\def\one{{\mathchoice {\rm 1\mskip-4mu l} {\rm 1\mskip-4mu l}
{\rm 1\mskip-4.5mu l} {\rm 1\mskip-5mu l}}}
\def\Q{{\mathchoice
{\setbox0=\hbox{$\displaystyle\rm Q$}\hbox{\raise 0.15\ht0\hbox to0pt
{\kern0.4\wd0\vrule height0.8\ht0\hss}\box0}}
{\setbox0=\hbox{$\textstyle\rm Q$}\hbox{\raise 0.15\ht0\hbox to0pt
{\kern0.4\wd0\vrule height0.8\ht0\hss}\box0}}
{\setbox0=\hbox{$\scriptstyle\rm Q$}\hbox{\raise 0.15\ht0\hbox to0pt
{\kern0.4\wd0\vrule height0.7\ht0\hss}\box0}}
{\setbox0=\hbox{$\scriptscriptstyle\rm Q$}\hbox{\raise 0.15\ht0\hbox to0pt
{\kern0.4\wd0\vrule height0.7\ht0\hss}\box0}}}}
\def\C{{\mathchoice
{\setbox0=\hbox{$\displaystyle\rm C$}\hbox{\hbox to0pt
{\kern0.4\wd0\vrule height0.9\ht0\hss}\box0}}
{\setbox0=\hbox{$\textstyle\rm C$}\hbox{\hbox to0pt
{\kern0.4\wd0\vrule height0.9\ht0\hss}\box0}}
{\setbox0=\hbox{$\scriptstyle\rm C$}\hbox{\hbox to0pt
{\kern0.4\wd0\vrule height0.9\ht0\hss}\box0}}
{\setbox0=\hbox{$\scriptscriptstyle\rm C$}\hbox{\hbox to0pt
{\kern0.4\wd0\vrule height0.9\ht0\hss}\box0}}}}

\font\fivesans=cmss10 at 4.61pt
\font\sevensans=cmss10 at 6.81pt
\font\tensans=cmss10
\newfam\sansfam
\textfont\sansfam=\tensans\scriptfont\sansfam=\sevensans\scriptscriptfont
\sansfam=\fivesans
\def\sans{\fam\sansfam\tensans}
\def\Z{{\mathchoice
{\hbox{$\sans\textstyle Z\kern-0.4em Z$}}
{\hbox{$\sans\textstyle Z\kern-0.4em Z$}}
{\hbox{$\sans\scriptstyle Z\kern-0.3em Z$}}
{\hbox{$\sans\scriptscriptstyle Z\kern-0.2em Z$}}}}

\newcount\foot
\foot=1
\def\note#1{\footnote{${}^{\number\foot}$}{\ftn #1}\advance\foot by 1}

\def\frac#1#2{{#1\over #2}}
\def\text#1{\quad{\hbox{#1}}\quad}

\font\ch=cmbx12 scaled\magstephalf
\font\ftn=cmr8 scaled\magstephalf
\font\ftit=cmti8 scaled\magstephalf
\font\it=cmti10 scaled\magstephalf

\font\titch=cmbx12 scaled\magstep2
\font\titname=cmr10 scaled\magstep2
\font\titit=cmti10 scaled\magstep1
\font\titbf=cmbx10 scaled\magstep2

\nopagenumbers

%%%%%%%%%%%%%%%%%%% title page %%%%%%%%%%%%%%%%%%%%%%%%%%%%%%%%%%

\line{\hfil }
\line{\hfil CGPG-94/5-1}
\line{\hfil May 10, 1994}
\vskip3cm
\centerline{\titch NEW LOOP REPRESENTATIONS FOR}
\vskip.5cm
\centerline{\titch 2+1 GRAVITY}
\vskip2cm
\centerline{\titname A. Ashtekar and R. Loll }
\vskip.5cm
\centerline{\titit Center for Gravitational Physics and Geometry}
\vskip.2cm
\centerline{\titit Physics Department, Pennsylvania State University}
\vskip.2cm
\centerline{\titit University Park, PA 16802, U.S.A.}

\vskip4cm
\centerline{\titbf Abstract}

Since the gauge group underlying 2+1-dimensional general relativity is
non-compact, certain difficulties arise in the passage from the
connection to the loop representations. It is shown that these
problems can be handled by appropriately choosing the measure that
features in the definition of the loop transform. Thus,
``old-fashioned'' loop representations -- based on ordinary loops -- do exist.
In the case when the spatial topology is that of a two-torus, these can
be constructed explicitly; {\it all} quantum states can be represented
as functions of (homotopy classes of) loops and the scalar product and
the action of the basic observables can be given directly in terms of
loops.

\vfill\eject
\footline={\hss\tenrm\folio\hss}
\pageno=1

%%%%%%%%%%%%%%%%%%%%%%%%%%%%%%%%%%%%%%%%%%%%%%%%%%%%%%%%%%%%%%%%%%%%%%%%%%

\line{\ch 1 Introduction\hfil}

Theories of connections are playing an increasingly important role in
the current description of all fundamental interactions in Nature.
The standard model of particle physics which encompasses the
electro-weak and strong interactions is based on Yang-Mills theories.
Classical general relativity (in three and four space-time dimensions) can
also be formulated as a dynamical theory of connections. Finally, such
theories are of interest from a mathematical viewpoint as well: many
of the recent advances in the understanding of the topology of
low-dimensional manifolds have come from theories of connections, in
particular from the analysis of Yang-Mills instantons and expectation
values of Wilson loop functionals in Chern-Simons theories.

In these theories, the configuration space generally arises as a space
$\cal A$ of connection one-forms $A(x)$ on a Cauchy surface $\Sigma$,
taking values in the Lie algebra of a gauge group $G$.  The
corresponding phase space $T^*\cal A$ is then naturally parametrized
by canonically conjugate pairs of fields $(A(x),E(x))$, where $E$ is a
vector density of weight one on $\Sigma$, taking values in the dual of
the Lie algebra of $G$, which can be thought of as a generalized
``electric" field conjugate to the gauge potential $A$.  Gauge
invariance is enforced by a Gauss constraint ${\rm Div}_A E (x) =0$.
As a consequence, the physical configuration space is
$\A/\G$, the quotient of $\A$ by the group $\G$ of local gauge
transformations. The physical observables are the gauge-invariant
functions on phase space.

Gambini and Trias [1] were the first to point out that a convenient
set of such observables can be associated with loops, i.e. closed
curves $\a$ in $\Sigma$ as follows. Choose a representation of
dimension $N$ of the gauge group $G$, and set
\vskip0.1cm
$$
\eqalign{&(T^0({\a}))\ [A]=\frac{1}{N}\, Tr\,\Big({\rm P}\exp \oint_\a
A \Big) \cr &(T^1 (\a,s))^a\ [A,E]= Tr\,\Big( E^a(\a (s))\,
{\rm P}\exp\oint_\a A \Big).}
\eqno(1.1)
$$
\vskip0.1cm
\ni where P denotes path ordering along the loop $\a$. $T^0(\a)$ is
labelled by the loop $\a$ and represents a (gauge-invariant)
configuration variable since it depends only on the connection $A$.
$(T^1(\a, s))^a$ is labelled by a loop $\a$ and a point $\a(s)$
on $\a$, and is a vector density at the point $\a(s)$.
Being linear in the electric field, it represents a
(gauge-invariant) momentum observable. These configuration and momentum
variables are closed under the Poisson bracket and constitute a
complete%
\note{Actually, these variables are {\ftit over}complete. This
occurs because the variables $T(\a)$ for different $\a$ are not all
independent; partly this is an expression of the fact that there are
``many more" loops $\a$ than points $x$ in $\Sigma$. In the algebraic
quantization method discussed below, the relations among these
variables have to be imposed in an appropriate fashion on the
quantum algebra.}%
set of functions on the phase space [2,3] (in the sense that their
gradients suffice to span the tangent space of the phase space almost
everywhere). Therefore, in an algebraic approach to quantization [3],
they can be chosen as the ``elementary variables'' which define the
``basic'' operators of the quantum theory. The task of quantization is
then reduced to that of finding appropriate representations of the
commutator algebra of these operators which mirror the Poisson
bracket algebra of $T^0(\a)$ and $(T^1(\a,s))^a$.

The obvious way to represent states is by suitable functionals $\Psi
[A]_{\G}$ of gauge equivalence classes of connections. This is the
configuration or the connection representation. The operators
$\hat{T}^0$ act by multiplication and the $\hat{T}^1$ by
(Lie-)derivation. Over the past three years, the mathematical problems
associated with these formal constructions have been analysed in
detail.  Specifically, integral and differential calculus has now been
developed on the space $\A/\G$ of connections modulo gauge
transformations, with the result that configuration representations
can now be constructed rigorously in the case when the underlying
gauge group is compact (for a summary, see, for example, [4]).

There is, however, another possibility: states can also be represented
as suitable functions of closed loops. This is
suggested by the possibility of making a ``Fourier-type'' transform
from the connection representation to a loop representation via
\vskip0.1cm
$$
\psi (\a):=\int_{{\cal A}/\cal G}d\mu([A]_{\cal G}) \ \
T^0_A(\a)[A]_{\cal G}\ \ \Psi [A]_{\cal G},\eqno(1.2)
$$
\vskip0.1cm
\ni where $\mu$ is a measure on $\A/\G$. This {\it loop transform} was
first introduced in the context of Yang-Mills theories by Gambini and
Trias [1] and later but independently in the context of general
relativity by Rovelli and Smolin [5]. In both cases, however, it was a
heuristic technique because the measure $\mu$ was not specified and
the required integration theory did not exist. Nonetheless, it has played a
powerful role as a heuristic device, especially in the context
of general relativity. In particular, it has suggested how one may translate
various operators acting on the connection states $\Psi[A]_{\G}$ to
operators on the loop states $\psi(\a)$. This in turn suggested how
to ``solve'' the quantum diffeomorphism constraint of general
relativity. In the loop representation, one can write down
the general solution to the diffeomorphism constraint as a loop
functional in the image of the transform, which depends only on the
(generalized) knot class to which the loop belongs.

Recent mathematical developments have made such considerations
rigorous in the case when the gauge group is compact. These results
can be summarized as follows. In the connection representation, states
are complex-valued functions on an appropriate completion $\agb$ of
$\ag$. This domain space of quantum wave functions, $\agb$, carries a
natural diffeomorphism-invariant measure $\mu_o$ which can be used to
rigorously define the loop transform (1.2). Operators such as
the $\hat{T}^0$ and the $\hat{T^1}$ can then be taken to the loop side
and used in various constructions.  Thus, in the case when the gauge
group is compact, loop representations exist rigorously.
\note{Note that each choice of measure defines a connection and
an equivalent loop representation. (In the connection representation,
the measure defines the inner product.) However, different choices of
measures $\mu$ on $\agb$ give rise to different representations. If
the systems under consideration have an infinite number of degrees of
freedom, the resulting representations are not generally expected to
be unitarily equivalent. It is the underlying physics that must determine the
appropriate measure and hence the appropriate representation.}

Loop representations have several appealing features. For example,
irrespective of the choice of the gauge group and the precise physics
contained in the theory, quantum states arise as suitably regular
functions on the loop space. That is, the domain space of quantum
states in various physical theories is the same. The regularity
conditions on loop states of course change from one theory to
another.  Nonetheless, since the domain space is ``unified'',
mathematical techniques can be shared between the various theories.
For diffeomorphism-invariant theories, one has the further advantage
that the action of the diffeomorphism group is coded more easily in
the loop space. In topological field theories where the connections
under consideration are all flat, one encounters an interesting
interplay between the quantum theory and the first homotopy group of
the manifold $\Sigma$. Finally, if $\Sigma$ is three-dimensional, one
has an avenue to explore knot invariants via theories of connections.

It is therefore natural to ask if the loop representations can also be
developed rigorously for physical theories -- such as general
relativity in three and four dimensions -- in which the gauge group $G$ is
non-compact.

A number of difficulties arises immediately. First, if the gauge group
is non-compact, the techniques [4] used to develop integration theory
over $\agb$ fail at a rather early stage. The problems here are not
insurmountable. However, a number of new ideas are needed and in
general the theory is likely to be considerably more complicated
unless, as in four-dimensional general relativity, the gauge group is
the complexification of a compact Lie group. It is therefore natural
to first restrict oneself to a context where these difficulties do not
arise. One such setting is provided by three-dimensional general
relativity. Here, the connections of interest turn out to be flat and
one can replace $\agb$ by the moduli space of flat connections.  Since
these spaces are finite-dimensional, one does not have to develop the
integration theory; the problems mentioned above do not arise. The
moduli space has several ``sectors''. On the sector where the traces
of holonomies $T^0_{\a}$ are all bounded, the transform can be defined
and the loop representation can be constructed in a straightforward
fashion [6].  However, it turns out that this sector does not
correspond to ``geometrodynamics''. On the sector which does, a new
difficulty arises: although the integration theory is straightforward,
the traces of holonomies $T^0_{\a}$ fail to be square-integrable with
respect to ``natural'' measures, making the loop transform analogous
to (1.2) ill-defined. Thus, on these physically interesting sectors, a
new strategy is needed.

This problem was first pointed out by Marolf [7] who also suggested a
way of tackling it in the special case where the manifold $\Sigma$ is
a torus $T^2$. The purpose of this paper is to suggest an alternative
solution, which consists of suitably modifying the measure that
appears in the transform. This solution is conceptually simpler in the
sense that with the new measures various problems are avoided right
from the beginning.  There is also a technical simplification. While
in the Marolf approach, one first restricts oneself to a suitable
dense subspace of the connection Hilbert space, defines the transform
and then extends it to the full Hilbert space, with the modified
measures, the transform exists on the full Hilbert space from the
outset.  More importantly, in our approach {\it the final result is a
genuine loop representation} in the sense that all states in the
Hilbert space are represented as functions of loops. By contrast, in
the Marolf approach the limiting states, which are not contained in
the initial dense subspace, cannot in general be represented as
functions of loops. This result had given rise to some concern about
the utility and the role of loops in the representation of quantum
states in the case when the gauge group is non-compact. Our analysis
clarifies this issue and shows that ``old-fashioned'' loop
representations, without the need of any smearing, do exist even in
the geometrodynamical sector of three-dimensional gravity. Our
solution does, however, have a drawback: our expressions for the
$\hat{T}^1$-operators in the loop representation are more complicated.
In spite of these differences, the final theory we obtain is unitarily
equivalent to Marolf's for $\Sigma =T^2$.  Therefore, the choice
between the two strategies is primarily a matter of taste and
convenience.

The outline of this paper is as follows. In Sec.2 we recall the basic
structure of 2+1 gravity on space-times $M=\Sigma^g\times\R$, with
$\Sigma^g$ a compact Riemann surface of genus $g$.  Sec.3 discusses
quantization in the connection representation and presents the general
strategy for modifying the measure to make the loop transform
well-defined. This strategy always leads to a ``regular'' loop
representation. In the case when the sector of the moduli space of
flat connections under consideration is compact and the traces of
holonomies $T^0_{\a}$ are bounded functions, the modification of the
measure is unnecessary.  However, if one chose to follow this route,
the resulting loop representation would be unitarily equivalent to
that of [6]. In the non-compact case, on the other hand, the strategy
appears to be essential to obtain a genuine loop representation. In
Sec.4, the procedure is carried out in detail for the case when
$\Sigma$ is a two-torus. In particular, we present a family of new
measures which make the loop transform well-defined and obtain the
modified expressions of the $\hat{T}^0$- and $\hat{T}^1$-operators as
well as the explicit form of the scalar product in the loop
representation for this family. Sec.5 contains our conclusions. In the
appendix we present some partial results for the genus-$2$ case.

\vskip1.5cm

\line{\ch 2  Preliminaries\hfil}

In this section, we will collect those results from the classical
Hamiltonian formulation of the 2+1-dimensional (Lorentzian) general
relativity that will be needed in the main part of the paper in Secs.3 and 4.
The discussion will also serve to fix our notation. Note that
this is not meant to be an exhaustive summary; further details may be
found, for example, in [3,8,9].

Since we are interested primarily in the canonical quantization of the
spatially compact case, we will restrict ourselves to
three-dimensional spacetimes $M$ of the form $M=\Sigma^g\times\R$,
where $\Sigma^g$ is a two-dimensional compact Riemann surface of genus
$g$. In the connection dynamics version, the Hamiltonian formulation
of the theory leads to two sets of first-class constraints [8,6,3],
one linear in momenta and the other independent of momenta.
Consequently, the Dirac and the reduced phase space quantization
methods lead to equivalent quantum theories. For convenience of
presentation, we will use the reduced phase space method here. Because
of the simplicity of the constraints, the reduced phase space is a
cotangent bundle over a reduced configuration space, which in turn is
just the moduli space ${\A}^F/{\G}$ of flat $SU(1,1)$-connections on
$\Sigma^g$. We will first recall relevant facts about ${\A}^F/{\G}$
and then go on to discuss the structure of the reduced phase space
$T^\star({\A}^F/{\G})$.

A flat $SU(1,1)$-connection $A$ on $\Sigma^g$ is determined by the
values of the $2g$ holonomies $U_i,\,i=1\dots 2g$, around
representatives $\a_i$ of the $2g$ homotopy generators
$\{\a_i\},\,i=1\dots 2g$, on $\Sigma^g$. Explicitly, we have
\vskip0.1cm
$$
U_i={\rm P}\exp{\oint_{a_i}A}.\eqno(2.1)
$$
\vskip0.1cm
\noindent Without loss of generality we will assume these representatives
to be based at a fixed point $p\in\Sigma^g$ and evaluate the
holonomies $U_i$ at $p$. Because of the fundamental relation between
the generators of the homotopy group, these holonomies are subject to
the condition
\vskip0.1cm
$$
U_1U_2U_1^{-1}U_2^{-1}U_3U_4U_3^{-1}U_4^{-1}\dots
U_{2g-1}U_{2g}U_{2g-1}^{-1}U_{2g}^{-1}=\one.\eqno(2.2)
$$
\vskip0.1cm
\noindent For computational purposes it is often useful to choose an
explicit parametrization for the $SU(1,1)$-matrices. We will set
\vskip0.1cm
$$
U_\alpha=\left( \matrix{\;\alpha_1+i\alpha_2&\alpha_3+i\alpha_4\cr
   \alpha_3-i\alpha_4&\alpha_1-i\alpha_2\cr}\right),\eqno(2.3)
$$
\vskip0.1cm
\noindent with real parameters $\alpha_1,...,\a_4$, subject to the
condition $\alpha_1^2+\alpha_2^2-\alpha_3^2-\alpha_4^2=1$. Our task is
to construct the moduli space of flat connections, i.e. to determine
the structure of the orbit space ${\cal A}^F/\cal G$. For this, we
note that $\cal G$ now acts on the holonomies $(U_1,...,U_{2g})$ at
the base point $p$ by the adjoint action of $U(p)\in SU(1,1)$
according to
\vskip0.1cm
$$
(U_1,U_2,\dots,U_{2g})\rightarrow
U(p)\cdot(U_1,U_2,\dots,U_{2g})\cdot U(p)^{-1}.\eqno(2.4)
$$
\vskip0.1cm
\noindent The moduli space is therefore obtained by incorporating (2.2)
and taking the quotient with respect to (2.4).

A key point for us is that the moduli space contains a finite number
of components, often referred to as ``sectors''. (For more precise
statements, see, for example, [9].)  This comes about because the isotropy
group $I$ of a holonomy $U_i$, i.e. the subgroup of $SU(1,1)$ leaving
fixed a particular $U_i$ under the adjoint action, is not universal
but depends on (certain properties of) $U_i$.  Let us associate a
3-vector $\vec\alpha_\perp:=(\alpha_2,\alpha_3,\alpha_4)$, with each
holonomy matrix $U_\alpha$, and define its norm as
\vskip0.1cm
$$
||\vec\alpha_\perp||:=\alpha_2^2-\alpha_3^2-\alpha_4^2. \eqno(2.5)
$$
\vskip0.1cm
\ni We can then distinguish the following cases%
\note{This classification is equivalent to that of [9] which is based
on the Lie algebra elements generating $U_i\in SU(1,1)$.}:
\item{i)} $||\vec\alpha_\perp||>0$, i.e. $||\vec\alpha_\perp||$ is
a timelike vector $\Longrightarrow I=SO(2)$,
e.g. for $\alpha=(1,0,0)$,
\vskip0.1cm
$$
I=\{ \left( \matrix{ e^{i\omega}&0\cr
0&e^{-i\omega}\cr } \right),\;\omega\in [0,2\pi]\};\eqno(2.6)
$$
\vskip0.1cm
\item{} A general timelike vector can be obtained by conjugation of
$U_\a$ with an arbitrary group element, in which case the isotropy group
$I$ likewise changes by conjugation. The same remark applies to the
cases below.
\item{ii)} $||\vec\alpha_\perp||<0$, i.e. $||\vec\alpha_\perp||$ is
a spacelike vector $\Longrightarrow I=\R\times \Z_2$, e.g. for
$\alpha=(0,1,0)$,
\vskip0.1cm
$$
I=\{ \left( \matrix{ \epsilon\cosh{\omega}&\sinh{\omega}\cr
\sinh{\omega}&
\epsilon\cosh{\omega}\cr } \right) ,\;\omega\in\R,\epsilon\in\Z_2\},\eqno(2.7)
$$
\vskip0.1cm
\item{} with group law $(\e_1,\omega_1)\cdot (\e_2,\omega_2)=(\e_1\e_2,
\e_1\omega_1+\e_2\omega_2)$;

\item{iii)} $||\vec\alpha_\perp||=0$, but not $\vec\a_\perp =0$, i.e.
$||\vec\alpha_\perp||$ is a non-vanishing null vector $\Longrightarrow
I=\R\times\Z_2$, e.g. for $\alpha=(1,1,0)$,
\vskip0.1cm
$$
I=\{ \left( \matrix{ \epsilon +i\omega &\omega\cr
\omega &\epsilon -i\omega\cr } \right) ,\;\omega\in\R,\epsilon\in\Z_2
\},\eqno(2.8)
$$
\vskip0.1cm
\item{} with group law $(\e_1,\omega_1)\cdot (\e_2,\omega_2)=(\e_1\e_2,
\e_1\omega_2+\e_2\omega_1)$. -- The isotropy group of the null vector
$\vec\alpha_\perp=0$ is of course the entire group.
\smallskip \noindent
In the sector that corresponds to geometrodynamics, all the $U_i$ are
boosts [10], corresponding to case ii) above. It turns out that this
sector is identical with the well-known Teichm\"uller space ${\cal
T}(\Sigma^g)$ associated with the Riemann surface $\Sigma^g$.  This is
a finite-dimensional space diffeomorphic to $\R^{6g-6}$ for $g>1$ and
$R^2$ for $g=1$. Thus, the geometrodynamic sector of the reduced
configuration space -- on which we will focus from now on -- is
precisely the Teichm\"uller space ${\cal T}(\Sigma^g)$.

Let us now turn to the reduced phase space $T^*{\cal T}(\Sigma^g)$.
We will give an explicit description of the moduli space of flat
connections in terms of gauge-invariant loop variables.

We begin by recalling [6,11] that there is an (over)complete set of
Dirac observables on the phase space $T^*({\cal A}^F/{\cal G})$, given by
the loop variables
\vskip0.1cm
$$
\eqalign{
T^0(\a)[A]=&\frac12 Tr\, U_{\a}\cr T^1(\a)[A,E]=&\oint_\a d\a^a
\,\eta_{ab}\,Tr\,( E^bU_{\a}),}\eqno(2.9)
$$
\vskip0.1cm
\ni where the canonical pairs $([A],[E])$ coordinatize the cotangent
bundle $T^*{\cal A}^F$ of the space ${\cal A}^F$ of flat connections,
$U_{\a} = {\rm P} \exp\oint_{\a} A$ is the holonomy around the loop
$\a$ evaluated at the base point, and where $\eta_{ab}$ is the totally
anti-symmetric Levi-Civita density on $\Sigma^g$. (Note that we have
exploited the fact that $\Sigma^{g}$ is two-dimensional to integrate
out $(T^1(\a, s))^a$ in (1.1) over the loop $\a$ to obtain a momentum
observable $T^1(\a)[A,E]$; the vector density index $a$ and the
dependence on the marked point $\a(s)$ are lost in the integration.)
Since all connections under consideration are flat, the variables
(2.10) depend only on the homotopy of the loop $\a$, and we may
substitute $\a$ by its corresponding homotopy group element.

The loop variables $T^I$, $I=0,1$, form a closed Poisson algebra with
respect to both the canonical symplectic structure on $T^*{\cal A}^F$,
and to the induced symplectic structure on the reduced, physical phase
space $T^*({\cal A}^F/{\cal G})$. The algebra is given by
\vskip0.1cm
$$
\eqalign{
&\{ T^0(\a),T^0(\b)\}=0\cr
&\{ T^0(\a),T^1(\b)\}=-\frac12\,\sum_i\Delta_i(\a,\b)\,
\Big(T^0(\a\circ_i\b)-T^0(\a\circ_i\b^{-1})\Big)\cr
&\{ T^1(\a),T^1(\b)\}=-\frac12\,\sum_i\Delta_i(\a,\b)\,
\Big(T^1(\a\circ_i\b)-T^1(\a\circ_i\b^{-1})\Big)\cr.}\eqno(2.10)
$$
\vskip0.1cm
\ni The sums in (2.10) are over all points $i$ of intersection of
the loops $\a$ and $\b$, with $\Delta_i(\a,\b)=1$ ($=-1$) if the two
tangent vectors $(\dot\a,\dot\b)$ form a right- (left-)handed dyad at
$i$ and zero if the tangent vectors are parallel. The algebra (2.10)
is independent of the representatives chosen in the homotopy classes
$\{\a\}$ and $\{\b\}$, and the representatives can be chosen to originate
and intersect at a fixed base point $p\in\Sigma$.  For this reason we
will from now on identify the loop composition $\circ_i$ with the
group multiplication $\circ$ in $\pi_1(\Sigma^g)$.

Because of the identities that hold among the traces of $2\times
2$-matrices, the $T^I$ are not all independent. They are subject to
the following algebraic relations:
\vskip0.1cm
$$
\eqalign{
T^0(\a) T^0(\b)=&\frac12\Big(T^0(\a\circ\b)+T^0(\a\circ\b^{-1})\Big)\cr
T^0(\a) T^1(\b)+T^0(\b)
T^1(\a)=&\frac12\Big(T^1(\a\circ\b)+T^1(\a\circ\b^{-1})\Big) . }
\eqno(2.11)
$$
\vskip0.1cm
\ni For a general gauge group, relations of this type are also known as
Mandelstam constraints. Finally, as an aside, note that the norm (2.5)
of $\vec\a_\perp$ is expressible in terms of the loop variable
$T^0(\a)$ introduced in (2.9) as
\vskip0.1cm
$$
||\vec\a_\perp||= \;1-T^0(\a)^2 ,\eqno(2.12)
$$
\vskip0.1cm
\ni which shows that the classification into timelike, spacelike and
null rotations made earlier is gauge-independent.

Since the reduced configuration space, the Teichm\"uller space ${\cal
T}$, is topologically trivial, one can attempt to find a global chart
on it using the $T^0$-functions. To achieve this, it is necessary to
eliminate the redundancy inherent in the Mandelstam constraints (2.11)
and to re-express and solve the condition (2.2) on holonomies as
conditions on the {\it traces} of the holonomies of the $2g$ homotopy
generators.

The overcompleteness of the $T^0$-variables has already been discussed
in a related case, namely that of $SU(2)$-holonomy variables of a
lattice gauge theory [11]. In so far as the arguments there were based
on the existence of constraints of the form (2.11) (which are the same
for both $SU(2)$ and $SU(1,1)$ in their two-dimensional
representations), they are equally valid in the present setting. Let
us summarize: given a set of $n$ basic loops $\a_i$ (here the $2g$
homotopy generators) and their associated holonomy matrices
$U_{\a_i}$, any gauge-invariant quantity $T^0(\g)$, where $\g$ is a
loop composed of the basic loops, can be expressed as an algebraic
function of the variables
\vskip0.1cm
$$
\eqalign{
&L_1(\a_i):= T^0(\a_i)\cr
&L_2(\a_i,\a_j):=\frac12 (T^0(\a_i\circ \a_j^{-1})-T^0(\a_i\circ\a_j)).}
\eqno(2.13)
$$
\vskip0.1cm
\ni This reduces the number of loop variables to $n+\frac{n(n-1)}{2}$. A
further reduction is provided by the following procedure. Fix two loops
(which can be thought of as ``projectors" in the Lie algebra [12]), say,
$\a_1$ and $\a_2$. Then any point in the space ${\cal A}/{\cal G}$ can be
described locally by the $3n-3$ variables
\vskip0.1cm
$$
\eqalign{
&L_1(\a_i),\;\; i=1,\dots,n \cr
&L_2(\a_1,\a_i),\;\;i=2,\dots,n \cr
&L_2(\a_2,\a_i),\;\;i=3,\dots,n.}\eqno(2.14)
$$

For $2+1$-gravity this leaves us with a set of $6g-3$ variables to describe
the space ${\cal A}^F/{\cal G}$. The fundamental relation (2.2) yields three
additional constraints on the variables (2.14), unless $g=1$, in which case
one obtains only one additional condition. Thus we end up with
$6g-6$ basic loop variables for $g\geq 2$ and $2$ basic loop variables for
$g=1$, coinciding with the dimensionality of the Teichm\"uller spaces.
Since moreover each space ${\cal T}(\Sigma^g)$ is contractible, there are
no obstructions in principle to finding sets of loop variables that
constitute a good global chart on it. Still those loop variables may
not independently assume arbitrary values on the real line, due to the
existence of inequalities among the variables (2.14) [13]. It is
fairly straightforward to explicitly identify the true physical degrees of
freedom in this manner. For general higher genus, one may follow the
strategy for the genus-$2$ case discussed in the appendix.

We will conclude this discussion by giving the general form of a
$SU(1,1)$-rotation $U$ in the space-like sector in terms of the
exponentiated connection components $A_i$, $i=2,3,4$, which will be
useful later (for simplicity, we consider only the part connected
to the identity,
where we have $T^0(\a)\geq 1$, $\forall\a$):
\vskip0.1cm
$$
\eqalign{
U_\a=\exp&\Big(
A_2\left( \matrix{i&0\cr 0&-i\cr}\right)+
A_3\left( \matrix{0&1\cr 1&0\cr}\right)+
A_4\left( \matrix{0&i\cr -i&0\cr}\right)\,\Big)\cr
 =&\left( \matrix{\cosh A+i\frac{A_2}{A}\sinh A&\frac{A_3+iA_4}{A}\sinh A\cr
   \frac{A_3-iA_4}{A}\sinh A&\cosh A-i\frac{A_2}{A}\sinh A}\right),}
\eqno(2.15)
$$
\vskip0.1cm
\ni where $A:=\sqrt{-A_2^2+A_3^2+A_4^2}$. Because of the condition
$||\vec\alpha_\perp||<0$, the square root is always well-defined. Given a
second holonomy matrix $U_\b$, obtained by exponentiating a connection with
components $B_i$, one can compute the explicit expressions for the loop
variables (2.14),
\vskip0.1cm
$$
\eqalign{
&L_1(\a)=\cosh A\cr
&L_1(\b)=\cosh B\cr
&L_2(\a,\b)=\frac{1}{AB}(A_2 B_2-A_3 B_3-A_4 B_4)\sinh A \sinh B.}
\eqno(2.16)
$$

\vskip1.5cm

\line{\ch 3 Quantization\hfil}

This section is divided into two parts. In the first, we recall
quantization in the connection representation. This discussion will
facilitate the introduction of the loop transform and also serve to
bring out some subleties. In the second, we first point out the
difficulties associated with the loop transform and then sketch our
proposal for overcoming them. In the next section, this strategy is carried out
in detail for the genus-one case .
\bigskip

\goodbreak
\ni {\it 3.1 Connection Representation}

By a quantization of $2+1$-gravity we shall mean a representation of
the algebra (2.10) as the commutator algebra
of self-adjoint operators $\hat{T}^I$ on some Hilbert space, such that
the conditions (2.11) hold (the products of operators on the left side
being replaced by their anti-commutators).

In the connection representation, we can proceed as one normally does
when dealing with quantum mechanics of systems whose configuration
space is a manifold. We can take the states to be densities
$\tilde\Psi[A]_{\G}$ of weight one-half on the reduced configuration
space ${\cal T}$, and let the scalar product be the obvious one:
\vskip0.1cm
$$
<\tilde\Psi|\tilde\Phi> := \int_{\cal T}\ \overline{\tilde\Psi}[A]_{\G}
\ \tilde\Phi[A]_{\G} ,
\eqno(3.1)
$$
\vskip0.1cm
\ni where a choice of a volume element on ${\cal T}$ is not necessary
because the integrand is a density of weight one. Since the $T^0$
are configuration variables, they are represented by multiplication
operators. Similarly, being momentum variables, the $T^1$
are represented by Lie derivatives,
\vskip0.1cm
$$
\eqalign{
&(\hat{T}^0_{\a}\circ \Psi) [A]_{\G} = T^0_{\a}[A]_{\G}\cdot
\Psi[A]_{\G}\cr
&(\hat{T}^1_{\a}\circ \Psi)
[A]_{\G} = {\hbar\over i}{\cal L}_{X(\a)} \Psi [A]_{\G},}\eqno(3.2)
$$
\vskip0.1cm
\ni where $X(\a)$ is the vector field on the reduced configuration space
${\cal T}$ which defines the classical momentum variable $T^1(\a)$.
(Recall that $T^1_{\a}$, being linear in momentum, is a contraction of
$[E]$ with a vector field on ${\cal T}$.  Thus, $X(\a)$ is the
projection to ${\cal T}$ of the vector field $\delta A$ on ${\cal
A}^F$, defined by $\delta A = \oint_{\a} d{\a}^a\,
\eta_{ab}\ \delta^2(x, \a(s))\ {\rm Tr}\, U_{\a}(s) (\delta/\delta A_b(x))$.)
It is straightforward to verify that with this definition the $\hat{T}^I_{\a}$
satisfy the canonical commutation relations that arise from (2.10)
and the algebraic conditions coming from (2.11).

In this description, the states naturally arose as scalar densities.
Given a volume element, one can ``de-densitize'' them and represent
them as the more familiar wave functions. This can be achieved using
{\it any} volume element $dV$ on ${\cal T}$. Furthermore, for any
$dV$, the action of the operators $\hat{T}^I$ can be translated in a
canonical fashion to $L^2({\cal T}, dV)$.

It turns out that the Teichm\"uller space ${\cal T}$ admits a
natural volume element.  To see this, we first note that there is a
natural symplectic structure $\Omega$ on the space of connections
$\cal A$ given by:
\vskip0.1cm
$$
\Omega(\d A,\d A')=\int_\Sigma d^2x\;\eta^{ab}(\d A)_a^i(\d
A')_{bi},\eqno(3.3)
$$
\vskip0.1cm
\ni for any two tangent vectors $\d A$, $\d A'$ to $\cal A$. This
form $\Omega$ can be pulled back to the space ${\cal A}^F$ of flat
connections. Being gauge-invariant, the pull-back in turn projects
down to Teichm\"uller space, where it coincides with the well-known
Weil-Petersson symplectic form%
\note{A global coordinate chart on Teichm\"uller space which is
convenient for our purposes is given by the Fenchel-Nielsen
coordinates (see, for example, [14]).  For $g>1$, they are a set of
length and angle coordinates $[l_i,\tau_i]$, $i=1\dots 3g-3$,
associated with a pants decomposition of the Riemann surface along a
set of $3g-3$ minimal geodesics. Moreover, they are canonical
coordinates for the symplectic form $\omega$, i.e. $\omega
=\sum_{i=1}^{3g-3}dl_i\wedge d\tau_i$.}
$\omega$.  This gives rise to a natural volume element $dV$ on $\cal
T$, namely, the $(3g-3)$-fold exterior product $dV_o=\omega\wedge\dots
\wedge\omega$. It is then natural to represent states as complex-valued
functions $\Psi[A]_{\G}$ and define the inner product as
\vskip0.1cm
$$
<\Psi|\Phi> := \int_{\cal T} dV_o \ \overline{\Psi}[A]_{\G}
\ \Phi[A]_{\G} .
\eqno(3.4)
$$

On this Hilbert space, the $\hat{T}^0_{\a}$ can be represented as before
by multiplication operators. They are densely defined and symmetric.
Normally, the definition of the momentum operators $\hat{T}^1(\a)$
would require a modification: if the Lie derivative of the volume
element with respect to the vector fields $X(\a)$ does not vanish, we
would have to add a multiple of the divergence of the vector field to
ensure that the resulting operator is symmetric. However, it turns out
[15] that the vector fields $X(\a)$ are in fact the Hamiltonian vector
fields on ${\cal A}^F$ for the symplectic structure $\omega$, where
the Hamiltonians are simply the functions $T^0_{\a}$ on $\cal T$.
Hence, in particular, the Lie derivative of the Liouville volume
element $dV_o$ with respect to any vector field $X(\a)$ vanishes.
Therefore, we can continue to represent $\hat{T}^1(\a)$ simply by the
Lie derivative. Thus, the representation of the basic operators is the
same as in (3.2), although the states are now wave functions
$\Psi[A]_{\G}$ on ${\cal T}$ rather than half-densities
$\tilde\Psi[A]_{\G}$. This formulation of the connection
representation will constitute the starting point for the discussion
of the loop transform in the next subsection.
\bigskip

\ni {\it 3.2 The loop transform}

We are now ready to construct the loop representation. The key idea
is to define this representation through a loop transform
of the type (1.2). In the present case, using the fact that the
reduced configuration space can be identified with the Teichm\"uller
space ${\cal T}$, we can simplify the transform to
\vskip0.1cm
$$
\psi(\a) = \int_{\cal T} dV\ T^0(\a)[l,\tau] \Psi[l,\tau] , \eqno(3.5)
$$
\vskip0.1cm
\ni where $dV$ is a volume element on ${\cal T}$ and we have used the
Fenchel-Nielsen coordinates [14] to parametrize $\cal T$ globally.
Thus, just as in the general context of Sec.1 the transform needed a
measure on $\agb$, the transform now requires the introduction of a
volume element on ${\cal T}$. Could we not have avoided the
introduction of this ad-hoc structure? After all, the transform has
the form of an inner product of $T^0(\a)$ with a wave function of
connections $\Psi$ and the connection representation {\it could} be
constructed intrinsically (i.e. without any additional structure such
as the volume element) if the states were represented by densities
$\tilde{\Psi}$ of weight one-half on ${\cal T}$.  Unfortunately, even
if we replaced the $\Psi[A]$ in (3.5) by $\tilde\Psi$, {\it because the
integral kernel of the transform,} $T^0(\a)$, {\it is a function} rather than a
density of weight one-half, we would still need an additional
structure (say, a fiducial density of weight one-half) to make the
integral well-defined.  Thus, while the connection representation
itself does not require the choice of a volume element, the passage to
the loop representation does%
\note{One might imagine defining the transform intrinsically
by using densities of weight one (rather than one-half) as
connection states. However, one would still need a volume element to
decide which of these densities of weight one are normalizable, i.e.
qualify to feature in the transform in the first place.}.

We saw at the end of Sec.3.1 that there {\it is} a natural
volume element $dV_o$ on ${\cal T}$ which arises from the
Weil-Petersson symplectic form. Therefore, a simple solution
to the problem would be to just choose this $dV_o$ for the required
volume element in (3.5). This strategy would work if we were
interested in the ``time-like'' sector of 2+1 gravity [6]. However, as
noted in Sec.2, in this paper we are interested in the ``space-like''
sector which corresponds to geometrodynamics. It is this choice that
led us to take the Teichm\"uller space ${\cal T}$ as the reduced
configuration space. Now, ${\cal T}$ is non-compact and in general
the $T^0(\a)$, being unbounded, fail to be square-integrable on $({\cal
T}, dV_o)$.  Hence, for a general quantum state $\Psi$ in the
connection representation -- which belongs to $L^2({\cal T}, dV_o)$ --
the integral in the transform would not be well-defined. In the
general setting considered here, this is the problem that was first
noted by Marolf [7] in the explicit context of the torus topology for
$\Sigma$.

A way out would be to first restrict the transform to a dense subspace
$D$ of $L^2({\cal T}, dV_o)$ -- such as the one spanned by the smooth wave
functions $\Psi [l,\tau]$ of compact support -- on which the transform {\it is}
well-defined, obtain the loop states and then take the Cauchy
completion of this space. While this procedure appears to be simple at
first sight, a detailed examination [7] shows that there are two key
problems. First, for the loop representation to exist, the dense space
$D$ has to satisfy {\it three} conditions: i) the integral on the
right of (3.5) must be well-defined for all $\Psi$ in $D$; ii) on $D$,
the transform should be faithful; and iii) $D$ should remain
invariant under the action of the $\hat{T}^I$-operators.
Although one does expect such
dense subspaces $D$ to exist, already in torus case it is a
quite non-trivial problem to find them. The second and conceptually
more important problem is that when one takes the Cauchy completion of
the image of $D$, one finds that it admits states which cannot, in a
natural way, be represented as functions of loops. Consequently, the
sense in which such a representation can be considered a ``loop
representation''  becomes rather obscure.

Our proposal therefore is to try a new strategy. The key idea is to
exploit the freedom in the choice of the volume element $dV$. Since we
are regarding $\cal T$ as a manifold, volume elements correspond to
$(6g-6)$-forms on $\cal T$. Hence, any two are related by a (suitably
regular) function. Thus, we can set $dV = m[l,\tau]dV_o$ for some
non-negative, smooth function $m[l,\tau]$. Following the terminology
common in physics, we will refer to $m[l, \tau]$ as a ``measure''. The
idea therefore is to choose an appropriately damped measure to make
the loop transform well-defined.

What conditions does $m[l,\tau]$ have to satisfy? First, as already noted, it
should ensure a sufficient damping so that the loop transform is well-defined.
More precisely, we will require that $m[l,\tau]$ be such that the traces of
holonomies $T^0(\a)$ are in $L^2({\cal T}, dV=m\, dV_o)$. Then, if we {\it
define} the connection representation using $dV$ -- which we are free to do --
we will be led to a well-defined loop transform. However, we also need the
action of the operators $\hat{T}^I(\a)$ on loop states to be well-defined and
manageable. For a general measure $m[l,\tau]$, the action of the
$\hat{T}^I(\a)$ reduces to
\vskip0.1cm
$$
\eqalign{
(\hat T^0(\a)\Psi) [l,\tau]&= T^0(\a)[l,\tau]\Psi [l,\tau]\cr
(\hat T^1(\a)\Psi) [l,\tau]&=-i\hbar( {\cal L}_{X(\a)}+\frac12
{\cal L}_{X(\a)}\ln m)\Psi [l,\tau],\cr}\eqno(3.6)
$$
\vskip0.1cm
\ni where, as before, ${\cal L}_{X(\a)}$ is the Lie derivative along
the vector field $X(\a)$ corresponding to the loop momentum observable
$T^1(\a)$, and where the second term on the right-hand side of the
relation for $\hat T^1$ compensates for the fact that the
Lie derivative of the volume element $dV= m\, dV_o$ with respect to
$X(\a)$ may not vanish.

These operator actions can be translated into the loop representation
via the transform (3.5). In order that the action of the resulting
operators be manageable -- so that in the final picture the loop
representation can exist in its own right -- it is necessary that the
term $\hat T^I(\b)T^0(\a)$ in
\vskip0.1cm
$$
\eqalign{
\Big(\hat T^I(\b)\psi\Big) (\a):=&\int_{\cal T} dV\,m[l,\tau]\,
T^0(\a) \Big(\hat T^I(\b)\Psi[l,\tau]\Big)\cr
 =&\int_{\cal T} dV\,m[l,\tau]\, \Big(\hat T^I(\b)T^0(\a) \Big)
\Psi[l,\tau]}\eqno(3.7)
$$
\vskip0.1cm
\ni be expressible as some linear combination $\sum a_jT^0(\a_j)$ (where
$\a_j$ denotes any homotopy group element). For $\hat T^I=\hat T^0$
this condition is automatically fulfilled, thanks to the algebraic
relation (2.11). For $I=1$, the term under consideration takes the
form
\vskip0.1cm
$$
({\cal L}_{X_\b}+\frac12 {\cal L}_{X_\b}\ln m)T^0(\a)=\{T^1(\b),T^0(\a)\}
+\frac12 \{T^1(\b),\ln m\} T^0(\a),\eqno(3.8)
$$
\vskip0.1cm
\ni where the curly brackets denote the Poisson brackets on phase space.
The first term on
the right-hand side is already in the required form, again due to the
algebraic relations (2.11). Next, it follows from (2.10) and (2.11)
that the second term, $\{T^1(\b),\ln m\}$, would have the desired form
if it were expressible as a linear combination of the $T^0$. Thus, the
action of $\hat{T}^1$ would be manageable on loop states if the
measure $m[l, \tau]$ were of the form
\vskip0.1cm
$$
m=\exp\Big( \sum_i b_i\,T^0(\a_i)\Big),\eqno(3.9)
$$
\vskip0.1cm
\ni for some fixed real constants $b_i$ and some fixed homotopy
generators $\a_i$.  (Note that we could also have chosen to use for
$m[l,\tau]$ the exponential of a product of $T^0(\a_i)$ since the
product can always be re-expressed as a sum, using (2.11).)

Our strategy is therefore to use a measure $m[l,\tau]$ of the form (3.9)
(both in the definition of the connection representation and) in the
definition of the loop transform. The key question is whether one can
choose a finite number of $b_i$ and $\a_i$ such that the measure damps
sufficiently fast for the transform (3.5) to be well-defined for
{\it any} element $\a$ of the homotopy group. At a heuristic level, it
would seem that the freedom in the choice of $\a_i$ is so large that
it should be easy to meet this damping condition. However, because we
do not have sufficient control over the behaviour of traces of
holonomies $T^0(\a)$ on the Teichm\"uller spaces of higher genus, we
have been able to explicitly demonstrate the existence of the measures
of the required type only in the $g=1$ case. However, if $b_i$ and
$\a_i$ can be chosen to ensure the existence of the integral in (3.5)
for all $\a$, the existence of a loop representation with the required
properties {\it is} ensured. In particular, all normalizable states in
such a representation would arise as functions of loops; in contrast
to [7], generalized loops would not be necessary.

\vskip1.5cm

\line{\ch 4  The torus case\hfil}

Let us briefly review the explicit structure of the reduced phase
space $T^*{\cal T}$ in the case when the two-manifold $\Sigma$ is a
torus.  (For further details, see [3,7].)  We are interested in the
sector where the holonomies $U_i$, $i=1,2$ of both homotopy generators
$(a_1,a_2)$ are rotations about spacelike axes in the three-dimensional
Minkowski space. The fundamental relation (2.2) implies that $U_1$ and
$U_2$ commute and are therefore rotations about the same axis. Without
loss of generality we may choose this axis to lie along the vector
$(0,1,0)$, which corresponds to setting $A_2=A_4=0$ in the holonomy
matrix (2.15). The reduced configuration space is therefore the
two-dimensional space $\cal T$ of flat connections on a torus $T^2$ and
can be parametrized by $\vec a\in \R^2$, with opposite signs
identified ($\vec a\sim -\vec a$), i.e. ${\cal T} =\R^2/\Z_2$.  The
corresponding reduced phase space is its cotangent bundle,
parametrized globally by the canonical variable pair $(\vec a,\vec
p)$.

The loop variables $T^i(\a)$ depend only on the homotopy class $\{\a\}$
of the loop $\a$, which for $\Sigma=T^2$ can be labelled by two
integers $\vec n$, characterizing the decomposition $\{\a\}=n_1\{\a_1\}
+n_2\{\a_2\}$.  The variables $T^0$ and their associated momentum
variables $T^1$ form an overcomplete set of observables on phase
space. Their explicit form is
\vskip0.1cm
$$
\eqalign{
&T^0(\vec k)[\vec{a}]=\cosh (\vec k\cdot\vec a)\cr
&T^1(\vec k) [\vec{a}, \vec{p}]=\sinh (\vec k\cdot \vec a)\,\vec k
\times\vec p,}
\eqno(4.1)
$$
\vskip0.1cm
\ni where $\vec k\times\vec p=k_1p_2-k_2 p_1$. Their Poisson
algebra can be written down explicitly:
\vskip0.1cm
$$
\eqalign{
&\{ T^0(\vec m),T^0(\vec n)\}=0\cr
&\{ T^0(\vec m),T^1(\vec n)\}=-\frac12\,( \vec m\times\vec n)\,
\Big(T^0(\vec m+\vec n)-T^0(\vec m-\vec n)\Big)\cr
&\{ T^1(\vec m),T^1(\vec n)\}=-\frac12\, (\vec m\times\vec n)\,
\Big(T^1(\vec m+\vec n)-T^1(\vec m-\vec n)\Big).}\eqno(4.2)
$$
\vskip0.1cm
\ni The overcompleteness of these variables is due to the Mandelstam
constraints (2.11), which now simplify to
\vskip0.1cm
$$
\eqalign{&
T^0(\vec m)T^0(\vec n)=\frac12\Big(T^0(\vec m+\vec n)
+T^0(\vec m-\vec n)\Big)\cr
&T^0(\vec m)T^1(\vec n)+ T^0(\vec n)T^1(\vec m)=
\frac12\Big(T^1(\vec m+\vec n)+T^1(\vec m-\vec n)\Big).}\eqno(4.3)
$$
\vskip0.1cm
\ni Note, as an aside,  that in the ``time-like'' sector where the moduli
space of connections is compact, the relevant loop observables are
obtained from (4.1) by substituting the hyperbolic functions with the
corresponding trigonometric functions, and the relation (4.2) and
(4.3) remain the same.

For the two generators of the homotopy group $\pi_1(T^2)$, $\a_1=(1,0)$ and
$\a_2=(0,1)$, one finds
\vskip0.1cm
$$
\eqalign{
&T^0(\a_1) [\vec{a}]=\cosh a_1\cr
&T^0(\a_2) [\vec{a}]=\cosh a_2\cr
&L_2(\a_1,\a_2)[\vec{a}] =-\sinh a_1 \sinh a_2.}\eqno(4.4)
$$
\vskip0.1cm
\ni It follows that $L_2(\a_1,\a_2)$ together with one of $T^0(\a_1)$,
$T^0(\a_2)$ parametrize $\cal T$ globally and would therefore
constitute a good choice of independent loop variables in terms of
which all other $T^0(\a)$ can be expressed. At first sight, the
easiest choice for an independent set may seem to take $T^0(\a_1)$
and $T^0(\a_2)$. However, they do not form a good global chart (a similar
statement holds for the genus-$2$ case, see the appendix).

In the coordinates $a_i$ on ${\cal T}$, the natural Liouville volume
element is simply $dV_o = da_1da_2$.  Our objective is to choose an
appropriate measure $m(a_i)=\exp -M(a_i)$ such that the transform and
the resulting loop representation are well-defined. To follow the
strategy outlined in Sec.3.2, let us begin by writing the analogs of
(3.5) -(3.9) explicitly. First, we have
\vskip0.1cm
$$
\psi(\vec n)=\;<T^0(\vec n),\Psi>\,=
\int_0^\infty da_1\,\int_0^\infty da_2\;e^{-M(\vec a)}\,
T^0(\vec n)\;\Psi(\vec a),\eqno(4.5)
$$
\vskip0.1cm
\ni using the scalar product notation. (Note that since the $T^0
(\vec{n})$ are all real, complex conjugation is unnecessary in the
integral.) According to our reasoning in Sec.3, the function $M$ on
${\cal T}$ should have the form $M = \sum b_i T^0(\a_i)$ (c.f. (3.9)),
where the $b_i$ and the $\a_i$ are such that holonomies
around arbitrary loops are square-integrable with respect to $dV
= (\exp -M) dV_o$. In the case of the torus, an obvious choice is
\vskip0.1cm
$$
M=c\,\Big( T^0(\vec q_1) +T^0(\vec q_2)\Big),\eqno(4.6)
$$
\vskip0.1cm
\ni where $c\in\R$, $c>0$, and $\vec q_1$ and $\vec q_2$ are linearly
independent homotopy classes. It turns out that for {\it any} positive
$c$ and any $\vec{q_1}$, $\vec{q}_2$, (4.6) leads to a loop representation
with all the desired properties discussed in Sec.3.
More precisely, we have the following:
\item{1.} It is straightforward to verify that every $T^0(\vec k)$ of
(4.1) belongs to $L^2({\cal T}, dV)$. Choosing this Hilbert space as
the space of states in the connection representation, it follows that
the transform, being simply the inner product, is a well-defined,
continuous map from the connection states to loop states. In
particular, while in the connection representation the states $\Psi$
are equivalence classes of functions on ${\cal T}$ (where two are
equivalent if they differ by a set of measure zero), their images
$\psi(\a)$ in the loop representation are genuine functions of
homotopy classes of loops.
\item{2.} The transform is {\it faithful}. To see this, note first that
for any choice of a measure $M$, $L^2({\cal T}, dV)$ provides an
irreducible representation of the algebra of operators $\hat{T}^I$,
defined by
\vskip0.1cm
$$
\eqalign{
&\hat T^0(\vec n)=\;\cosh (\vec n\cdot\vec a)\cr
&\hat T^1(\vec n)=-i\hbar\sinh (\vec n\cdot\vec a)\;\vec n\times\Big(
\frac{\del}{\del \vec a}+\frac12\vec\nabla M\Big).} \eqno(4.7)
$$
\vskip0.1cm
\item{} In particular, this is true for our choice (4.6).  Now,
suppose the transform has a kernel ${\cal K}$. Then, {\cal K} is a
closed subspace of $L^2({\cal T}, dV)$ and, since (4.6) is of the type
(3.9), it follows that ${\cal K}$ remains invariant under the action
of the $\hat{T}^I$. Hence, ${\cal K}$ must be either the zero subspace
or the full Hilbert space. It cannot be the full Hilbert space
because, in particular, the elements $T^0(\a)$ of $L^2({\cal T}, dV)$
cannot lie in the kernel for any $\a$ since its norm is positive
definite. Hence ${\cal K}$ must contain only the zero vector.
\item{3.} It follows from our discussion of the connection representation that
for any positive $M$, the representation (4.7) of the $\hat{T}^I$-algebra
on $L^2({\cal T}, dV= (\exp -M) dV_o)$ is unitarily equivalent
to the representation on $L^2({\cal T}, dV_o)$. Hence, in particular,
it follows that the loop representations of the $\hat{T}^I$-algebra
obtained by using different $c\ge 0$, and independent ${\vec q_1}$,
$\vec{q_2}$ are also unitarily equivalent. In particular then, these
``genuine'' loop representations are unitarily equivalent to the one
constructed by Marolf [7].

The transform enables us to endow the loop states with an inner product
and to represent the observables ${\hat T}^I$ directly on
the loop states.
The action (4.7) of the $\hat{T}^I(\a)$-operators translates to the
loop representation yielding
\vskip0.1cm
$$
\eqalign{
&\Big(\hat T^0(\vec k)\psi\Big)
(\vec n)=\frac12\Big( \psi(\vec n +\vec k)
+\psi(\vec n -\vec k)\Big)\cr
&\Big(\hat T^1(\vec k)\psi\Big)
(\vec n)=-\frac{i\hbar}{2}(\vec k\times\vec n)\Big(
\psi(\vec k +\vec n)-\psi(\vec k-\vec n)\Big)+\cr
\quad & +\frac{ic\hbar}{4}\sum_{i=1,2}
 (\vec k\times\vec q_i)\Big(\psi(\vec k+\vec n+\vec q_i)-
\psi(\vec k+\vec n -\vec q_i)+\psi(\vec k -\vec n +\vec q_i)-
\psi(\vec k-\vec n -\vec q_i)\Big).}\eqno(4.8)
$$
\vskip0.1cm
\ni Using these expressions, one may check explicitly
that the commutation relations that result from (2.10) and the
algebraic conditions that arise from (2.11) are satisfied by the
$\hat{T}^I$ in this loop representation.

Finally, let us exhibit the inner products between loop states.  For
simplicity, let us consider the subset of loop states $\{T^0_A(\vec
k),\, \vec k\in \R^2/\Z_2\}\subset L^2({\cal T},m\,dV_o)$, and, in the
measure, set $\vec q_1=(1,0)$ and $\vec q_2=(0,1)$.  The general
expression for the scalar product is then
\vskip0.1cm
$$
\eqalign{
<T^0(\vec k),T^0(\vec n)>=&2\int_0^\infty da_1\int_0^\infty da_2 \,
e^{-c(\cosh a_1 +\cosh a_2)} T^0(\vec k)T^0(\vec n)\cr
=\,&F(k_1+n_1) F(k_2+n_2) + F(k_1-n_1) F(k_2-n_2) +\cr
 +&G(k_1+n_1) G(k_2+n_2) + G(k_1-n_1) G(k_2-n_2),}\eqno(4.9)
$$
\vskip0.1cm
\ni where
\vskip0.1cm
$$
\eqalign{
F(0&)=K_0(c)\cr
F(1&)=K_1(c)\cr
F(n&)=K_0(c)\,\Big( 2^{n-1} +n\sum_{k=1}^{\frac{n}{2}} (-1)^k \frac{1}{k}
\left({n-k-1\atop k-1}\right) 2^{n-2k-1}\Big)+
    \sum_{m=1}^{\frac{n}{2}}(2m-1)!! \frac{1}{c^m} K_m(c)\times\cr
\times\Big( &2^{n-1}\left( {\frac{n}{2}\atop \frac{n}{2}-m} \right) +
n \sum_{k=1}^{\frac{n}{2}-m} (-1)^k \frac{1}{k} \left({n-k-1 \atop k-1}
\right) 2^{n-2k-1} \left( {\frac{n}{2}-k\atop \frac{n}{2}-k-m}\right),\;
n\,{\rm even}\cr
&\cr
F(n&)=K_1(c)\,\Big( 2^{n-1} +n\sum_{k=1}^{\frac{n-1}{2}} (-1)^k
\frac{1}{k}
\left({n-k-1\atop k-1}\right) 2^{n-2k-1}\Big)+
\sum_{m=1}^{\frac{n-1}{2}}(2m-1)!! \frac{1}{c^m} K_{m+1}(c)\times\cr
\times\Big( &2^{n-1}\left( {\frac{n-1}{2}\atop \frac{n-1}{2}-m} \right) +
n \sum_{k=1}^{\frac{n-1}{2}-m} (-1)^k \frac{1}{k} \left({n-k-1 \atop
k-1}\right) 2^{n-2k-1} \left( {\frac{n-1}{2}-k\atop
\frac{n-1}{2}-k-m}\right),\; n\,{\rm odd}\cr
&\cr
G(0&)=0\cr
G(n&)=e^{-c}\sum_{m=1}^n \frac{1}{c^m}\sum_{k=0}^{E(\frac{n-m}{2})} (-1)^k
\left({n-k-1\atop k}\right) 2^{n-2k-1}\frac{(n-2k-1)!}{(n-2k-m)!}.
}\eqno(4.10)
$$
\vskip0.1cm
\ni Here, the functions $K_n$ are the modified Bessel functions of
integer order [16]. It is important to notice that, unlike in
the case of the ``time-like sector'' [6], the scalar product is {\it
not} proportional to $\d_{\vec k,\vec n}$. Note also that, as
expected, the above expressions diverge as $c\rightarrow 0$
(corresponding to the trivial measure $M=0$), due to the divergence of
the Bessel functions and of the terms proportional to $c^{-m}$ in
(4.14). Let us give a few special cases of (4.13) for illustration:
\vskip0.1cm
$$
\eqalign{
<T^0(0,0),T^0(0,0)>=&2K_0(c)^2\cr
<T^0(1,0),T^0(1,0)>=&<T^0(0,1),T^0(0,1)>=2K_0(c) (\frac{1}{c}K_1(c)+K_0(c))\cr
<T^0(1,0),T^0(0,0)>=&<T^0(0,1),T^0(0,0)>=2K_0(c)K_1(c)\cr
<T^0(1,0),T^0(0,1)>=&2K_1(c)^2\cr
\vdots}\eqno(4.11)
$$

Finally we note that, although the wave functions $T^0(k,n)$ do not
form an orthonormal set, since the weight function $e^{-M(\vec
a)}$ is everywhere positive on $\cal T$, there exists a sequence of
polynomials $P_m(a_1)Q_n(a_2)$, where
\vskip0.1cm
$$
\eqalign{
&P_m(a_1)=\sum_{k=1}^m P_{mk}T^0(k,0)\cr
&Q_m(a_2)=\sum_{k=1}^m Q_{mk}T^0(0,k),}\eqno(4.12)
$$
\vskip0.1cm
\ni with constants $P_{mk},\;Q_{mk}$, such that
\vskip0.1cm
$$
<P_kQ_l,P_mQ_n>=\delta_{km}\delta_{ln}.\eqno(4.13)
$$
\vskip0.1cm
\ni Such a basis would still have a discrete labelling, but the direct
geometric interpretation of the wave functions in terms of loops on
$T^2$ would be lost.

One may object to the introduction of a non-trivial measure factor $\exp -M$ on
the grounds that the trivial measure $da_1\wedge da_2$ is distinguished by its
modular invariance, i.e. invariance under the action of the modular group,
whose generators act on the connection variables $(a_1,a_2)$ according to
\vskip0.1cm
$$
\eqalign{
&(a_1,a_2)\rightarrow (a_2, -a_1)\cr
&(a_1,a_2)\rightarrow (a_1, a_1+a_2).}\eqno(4.14)
$$
\vskip0.1cm
\ni Modular
invariance (i.e. invariance under large diffeomorphisms) however is not a
physical requirement of the 2+1-theory, and its imposition leads to orbifold
singularities in the reduced configuration space [17]. Although our
modified measures are not modular invariant, their corresponding quantum
representations still allow for a unitary implementation of the modular group.
The action of the generators on wave functions is given by
\vskip0.1cm
$$
\eqalign{
&\Psi (a_1,a_2)\rightarrow e^{(M(a_1,a_2)-M(a_2,-a_1))/2}\,
\Psi(a_2,-a_1)\cr
&\Psi (a_1,a_2)\rightarrow e^{(M(a_1,a_2)-M(a_1,a_1+a_2))/2}\,
\Psi (a_1,a_1+a_2).}\eqno(4.15)
$$
\vskip0.1cm
\ni For our particular choice $M=c (T^0(1,0)+T^0(0,1))$ one obtains
\vskip0.1cm
$$
\eqalign{
&\Psi (a_1,a_2)\rightarrow \Psi (a_2,-a_1)\cr
&\Psi (a_1,a_2)\rightarrow e^{\frac{c}{2}
(T^0(0,1)-T^0(1,1))}\,\Psi(a_1,a_1+a_2).}\eqno(4.16)
$$
\vskip0.1cm
\ni This action takes on a simple form in the loop representation only if one
makes a change of basis (to one in which the operators $\hat T^0(\vec k)$
are diagonal).

\vskip1.5cm

\line{\ch 5 Conclusion\hfil}

Let us begin with a summary of the main results. While the connection
representation can be constructed without a volume element or a
measure on ${\cal T}$, the loop transform does require this additional
structure.  The Teichm\"uller space ${\cal T}$ is equipped with a
natural, Liouville volume element $dV_o$.  Unfortunately, because the
traces of holonomies grow unboundedly in the coordinates which are
canonical for the Liouville form, the integral in the loop transform
defined using $dV_o$ is in general ill-defined. Our strategy was to
exploit the freedom available in the choice of the volume element to
introduce a damping factor $\exp -M$ ($M\ge 0$) and define the
transform using $dV= (\exp -M) dV_o$ instead.  The transform is then
well-defined if $M$ is chosen so that all the traces of holonomies
$T^0_{\a}$ belong to $L^2({\cal T}, dV)$.  The requirement that the
observables $\hat{T}^I$ have manageable expressions in the loop
representation further restricts the damping factor $M$: it has to be
of the type $\sum b_i T(\a^o_i)$ for some real numbers $b_i$ and
homotopy generators $\{\a^o_i\}$. The key question then is whether the
two requirements on the damping factor can be met simultaneously. In
the genus $g=1$ case, we saw that it was quite straightforward to
achieve this. In the more general case, the issue remains open
although the available freedom in the choice of
constants and homotopy generators seems large enough to meet
these conditions.

The solution we propose here does have an inelegant feature: the
expressions of the operators $\hat{T}^1(\a)$ in the loop
representation now involve not only the homotopy generator $\a$
labelling the operator but also the fiducial loops $\a^o_i$ we fixed
to define the measure. Could we have avoided this by modifying the
strategy slightly? For example, in the definition of the transform,
$\psi(\a) = \int (\exp -M) dV_o T^0(\a)\Psi$, could we not have
constructed the damping factor $M$ from the homotopy generator $\a$
itself, without introducing any fiducial generators $\a^o_i$? This is
a tempting strategy since it avoids all references to fiducial loops.
However, it does not work, essentially because the transform no longer
has the form of an inner product of $\Psi$ with $T^0(\a)$. More
specifically, in the resulting loop representation, it is not possible
to express the action of even the $\hat{T}^0$ operators in a
manageable way!

A second strategy [5] would be to avoid the loop transform altogether
and introduce the loop representation ab-initio. Thus, one may begin
with the quantum algebra of the $\hat{T}^I$-operators and attempt to find
a representation directly on a vector space of suitable
functions $\psi(\a)$ of (homotopy classes of) loops.  Unfortunately,
any ansatz which avoids the introduction of the transform and
reference to the connection representation faces two important
problems.  First, in such representations, it seems difficult to
incorporate the numerous identities and inequalities satisfied by the
loop variables [13].  Second,  it is difficult to simply guess the
class of ``suitable'' functions $\psi(\a)$ one has to begin with.
In the case of 2+1 gravity on a torus, for example, it would a priori
seem natural to begin with functions $\psi(\vec{n}) = \sum c_i
\delta_{\vec n, \vec n_i}$, obtained by taking linear combinations of
characteristic functions of homotopy classes. When the quantization
program [3] is completed, however, one finds that the spectrum of
all the $\hat{T}^0({\vec n})$-operators is bounded between $(-1, 1)$
while classically, on the geometrodynamic sector, the classical
$T^0(\a)({\vec n})$ take values precisely in the complement of this
interval!  That is, harmless assumptions on the initial ``regularity''
conditions end up having unexpected, physically important and often
undesired consequences. In the example just described, the quantum
theory can be constructed but it corresponds to the ``time-like
sector'' of the moduli space of flat connections which has no
geometrodynamic analog.  More importantly, the ``correct'' regularity
conditions that will finally lead one to the desired sector may be
quite involved and difficult to guess because one's favourite loop
states -- such as the characteristic functions of homotopy classes --
may not belong to the physical Hilbert space.  This is the case both
for the representations presented in the last section and the ones
found by Marolf [7]. All these problems are avoided if one constructs
the loop representation via the loop transform.

2+1 gravity is a ``toy model'' for the physical 3+1 theory.  What
lessons can one learn from its analysis? First, we
found that, once appropriate care is taken in defining measures, ``old
fashioned'' loop representations {\it do} exist for the 2+1 theory.
Results of [7] had been used by some to question the existence and
utility of the loop representation in cases when the gauge group is
non-compact. Our analysis removes these objections. It does point out,
however, that even in absence of infinite-dimensional, field-theoretic
problems the task of choosing physically interesting sectors can be
a rather delicate issue in loop representations. In the
connection representation of the 2+1 theory, restricting
oneself to the geometrodynamic sector was straightforward: we simply
restricted the wave functions to have support in this sector. In the
loop representation, by contrast, the restriction is implemented by
imposing different ``regularity conditions'' on loop states which, in
turn, lead to quite different inner products. Without recourse to the
transform, it would have been hopeless to unravel this subtle
intertwining between the physics of the representation and the
mathematical regularity conditions. The second lesson therefore is that it
would be ``safer'' to construct the loop representation via the
transform also in the 3+1 theory. The relation between the mathematical
assumptions and their physical implications would then be more
transparent and the numerous identities and inequalities [13] between
traces of holonomies automatically incorporated in the loop
representation.  This in turn suggests that the construction of
suitable measures on $\ag$ would be a central problem in the 3+1 theory as
well. This is the third lesson. The details of the required strategies
in 3+1 dimensions will, however, be quite different from those that
have been successful in the present analysis. The 3+1-problem is both
more difficult and easier.  It is more difficult because the space
$\ag$ is now {\it infinite}-dimensional. However, it is also easier
because the gauge group $\C SU(2)$ of the 3+1 theory is the
complexification of a {\it compact} group, and states in the connection
representation are {\it holomorphic} functions of connections. It
therefore seems possible that the integration theory on $\ag$
developed for compact groups [4] would admit a natural generalization to
this case. This approach would lead to measures of a quite different
sort than the ones introduced in this paper and in particular would not
refer to any fiducial loops.

\vskip1.2cm
\ni{\it Acknowledgements:} We would like to thank Jorma Louko and Don
Marolf for discussions and Steve Carlip for pointing out reference [14]. This
work was supported in part by the NSF grant PHY93-96246, the Eberly
research fund of Penn State University, and a DFG fellowship to RL.

\vskip1.5cm
\line{\ch Appendix\hfil}

In this appendix we present some details on the genus-2 case. In
particular, we will demonstrate that the problems encountered in the genus-1
case persist and therefore again a measure with an appropriate
damping factor is needed.

The reduced configuration space for the genus-2 case is the
six-dimensional Teichm\"uller space and may be parametrized globally
by the Fenchel-Nielsen coordinates [14].  They are a set of length and
angle coordinates of a pants decomposition of the genus-g surface. The
surface is cut along a set of $3g$ simple geodesic loops and, at the
i'th cut, $l_i\in \R^+$ measures the intrinsic length of the border
curve and $\tau_i\in \R$ the relative twisting angle of the opposite
sides of the cut. As was shown by Wolpert [18], the Weil-Petersson
symplectic form $\omega$ in these coordinates is simply given by
$\omega =\sum_{i=1}^{3g-3}dl_i\wedge d\tau_i$.

To find a suitable damping factor, we need estimates on how fast the
traces of holonomies diverge on ${\cal T}$. Thus, we have to express
the loop holonomies as functions of the $[l_i,\tau_i]$.
For $g\geq 2$ this task has been carried out by Okai [19].

We will adopt the notation of [19] and denote the six Fenchel-Nielsen
parameters by $(l_{-\infty}, l_0, l_\infty, \tau_{-\infty}, \tau_0,
\tau_\infty)$.  Next one has to choose a set of six loops on $\Sigma$
such that the traces of their holonomies are independent, i.e.
parametrize the Teichm\"uller space locally. There are clearly many
different ways of doing this. The simplest choice we have found is
illustrated in Fig.1:

%\vskip5cm
\vskip1cm
\epsffile{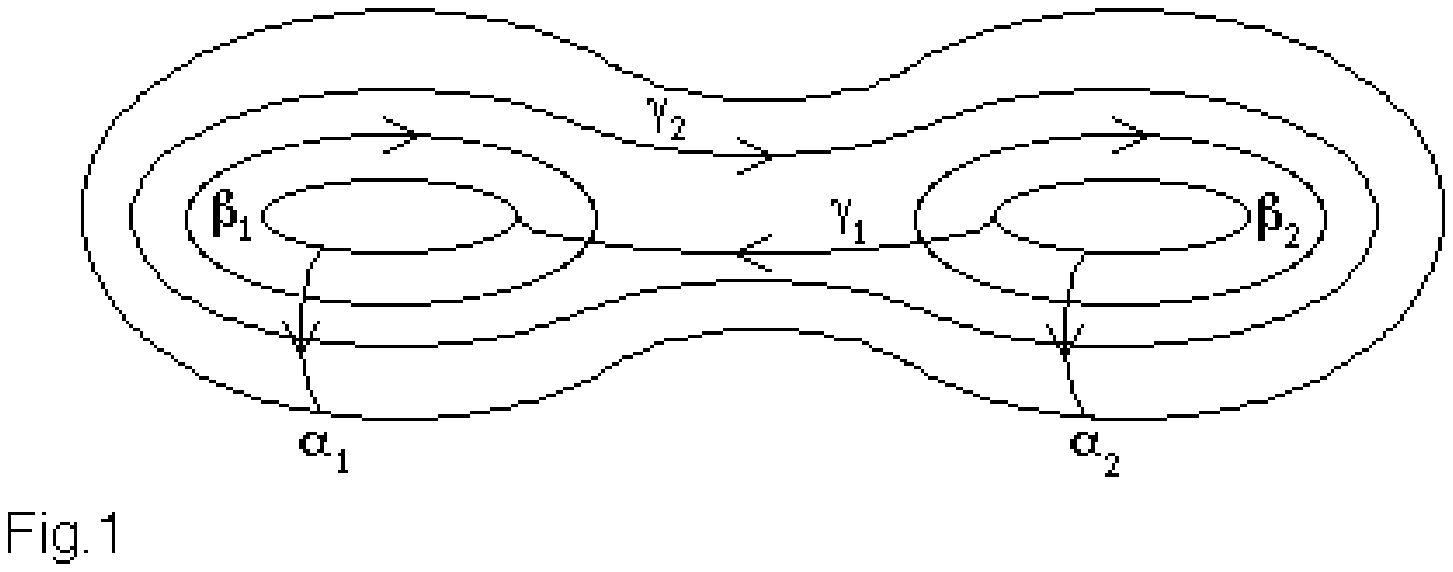}

\ni The loops $(\alpha_1, \alpha_2, \beta_1, \beta_2)$ are the usual
homotopy generators, and in addition we have the two loops
$(\gamma_1,\gamma_2)$. The normalized traced holonomies of these loops
are the following functions of $(l_i,\tau_i)$:
\vskip0.1cm
$$
\eqalign{
L_1(\alpha_1) &= \cosh{\frac{l_{-\infty}}{2}}\cr
L_1(\alpha_2) &= \cosh{\frac{l_{\infty}}{2}}\cr
L_1(\gamma_1) &= \cosh{\frac{l_0}{2}}\cr
L_1(\beta_1) &= \sinh{\frac{\tau_{-\infty}}{2}}\sinh{\frac{\tau_0}{2}} +
  \cosh{s_\infty} \cosh{\frac{\tau_{-\infty}}{2}}\cosh{\frac{\tau_0}{2}}\cr
L_1(\beta_2) &= \sinh{\frac{\tau_{\infty}}{2}}\sinh{\frac{\tau_0}{2}} +
  \cosh{s_{-\infty}} \cosh{\frac{\tau_{\infty}}{2}}\cosh{\frac{\tau_0}{2}}\cr
L_1(\gamma_2) &= \sinh{\frac{\tau_{-\infty}}{2}}\sinh{\frac{\tau_\infty}{2}} +
  \cosh{s_0} \cosh{\frac{\tau_{-\infty}}{2}}\cosh{\frac{\tau_\infty}{2}},}
\eqno(A.1)
$$
\vskip0.1cm
\ni where the length parameters $s_{-\infty}$, $s_0$ and $s_\infty$ are
functions of the $l_i$ alone,
\vskip0.1cm
$$
\eqalign{
&\cosh{s_{-\infty}}=\frac{\cosh{\frac{l_\infty}{2}}\cosh{\frac{l_0}{2}} +
  \cosh{\frac{l_{-\infty}}{2}} }{\sinh{\frac{l_\infty}{2}}\sinh{\frac{l_0}{2}}}
\cr
&\cosh{s_0}=\frac{\cosh{\frac{l_\infty}{2}}\cosh{\frac{l_{-\infty}}{2}} +
  \cosh{\frac{l_0}{2}} }{\sinh{\frac{l_\infty}{2}}\sinh{\frac{l_{-\infty}}{2}}}
\cr
&\cosh{s_{\infty}}=\frac{\cosh{\frac{l_{-\infty}}{2}}\cosh{\frac{l_0}{2}} +
\cosh{\frac{l_\infty}{2}}}{\sinh{\frac{l_{-\infty}}{2}}\sinh{\frac{l_0}{2}}}.}
\eqno(A.2)
$$

Note that the particular $L_1$'s chosen in (A.1), unlike the
Fenchel-Nielsen variables, are not good global coordinates on
Teichm\"uller space, since a simultaneous sign change of the $\tau_i$
leads to the same values for the independent $L_1$-variables. It is
obvious from (A.1) and (A.2) that a problem similar to that
encountered in the genus-1 case arises here too (and, in fact, for any
higher genus), since none of the loop variables in (A.1) are
square-integrable with respect to the measure $\Pi_i\,dl_i d\tau_i$.
Moreover, there is now an additional problem for small $l$, namely,
another divergence in the loop transform coming from terms like
$\cosh{s_\infty}$. This problem also occurs in some calculations in
string theory (see, for example, [20]), and may be dealt with by
introducing a cut-off for small lengths.

Unfortunately, it appears difficult to extract from (A.1) and (A.2)
estimates for the asymptotic growth of traces of holonomies around
arbitrary elements of the homotopy group. This happens because --
although possible in principle -- it is in practice difficult to
express arbitrary $T^0 (\a)$ as functions of the independent set.
Thus, while it is tempting to choose the damping factor simply as
\vskip0.1cm
$$
\exp -\sum_{i=1}^2 [L_1(\a_i) +L_1(\b_i)+ L_1(\g_i)],\eqno(A.3)
$$
\vskip0.1cm
\ni there is no guarantee that this damping will suffice to make the loop
transform well-defined for any $\a$ in the homotopy group.

Even if one restricts the quantum wave functions to sums of tensor
products of the six basic loop functions $L_1(\a_1), ..., L_1(\g_2)$,
the situation is still non-trivial.  Since some of those functions
themselves have a complicated, coupled dependence on the
Fenchel-Nielsen parameters, it is not immediately clear whether (A.3)
gives a sufficient damping in all of the asymptotic regions.  One
could probably get more control over the measure by re-expressing it
in terms of the independent variables $L_1$. The Jacobian of this
transformation can be readily expressed as a simple function of
$L_2$-variables (c.f. (2.13)), but again there is no straightforward
way of rewriting it in terms of the independent set $\{L_1\}$.  The
algebraic problems encountered here are not insurmountable, however, a
detailed case by case analysis appears to be necessary to obtain a
complete control on the asymptotic behaviour of the traces of holonomies
around arbitrary loops. Only then will one be able to
construct explicit measures that make the loop transforms for
higher-genus surfaces well-defined.

\vskip2cm

\vfill\eject

\line{\ch References\hfil}

\item{[1]} Gambini, R. and Trias, A.: Gauge dynamics in the C-representation,
  {\it Nucl. Phys.} B278 (1986) 436-448
\item{[2]} Goldberg, J., Lewandowski, J. and Stornaiolo, C.: Degeneracy in
  loop variables, {\it Commun. Math. Phys.} (1992) 148, 377; Ashtekar, A.
  and Lewandowski, J.: Completeness
  of Wilson loop functionals on the moduli space of $SL(2,\C)$ and $SU(1,1)$
  connections, {\it Class. Quant. Grav.} (1993) L69-74.
\item{[3]} Ashtekar, A. {\it Lectures on Non-Perturbative Canonical Gravity},
  World Scientific, Singapore, 1991
\item{[4]} Ashtekar, A.,  Marolf, D. and Mour\~ao, J.: Integration on the
  space of connections modulo gauge transformations, in: {\it Proceedings of
  the Lanczos centenary conference}, ed. D. Brown et al., SIAM
\item{[5]} Rovelli, C. and Smolin, L.: Loop space representation of quantum
  general relativity, {\it Nucl. Phys.} B331 (1990) 80-152
\item{[6]} Ashtekar, A., Husain, V., Rovelli, C., Samuel, J., and Smolin, L.:
  2+1 gravity as a toy model for the 3+1 theory, {\it Class. Quant. Grav.}
  6 (1989) L185-193; Smolin, L.: Loop representation for quantum gravity in
  2+1 dimensions, in: {\it Knots, topology and quantum field theory},
  Proceedings of the 12th Johns Hopkins Workshop; Ashtekar, A.: Lessons from
  2+1 dimensional quantum gravity, in: {\it Strings 90}, ed. R. Arnowitt et
  al., World Scientific, Singapore
\item{[7]} Marolf, D.M.: Loop representations for 2+1 gravity on a torus,
  {\it Class. Quant. Grav.} 10 (1993) 2625-2647
\item{[8]} Witten, E.: 2+1 dimensional gravity as an exactly soluble system,
  {\it Nucl. Phys.} B311 (1989) 46-78
\item{[9]} Louko, J. and Marolf, D.M.: The solution space of 2+1-gravity
  on $\R\times T^2$ in Witten's connection formulation, {\it Class.
  Quant. Grav.} 11 (1994) 311-330;
  Carlip, S.: Six ways to quantize (2+1)-dimensional gravity,
  to appear in {\it Proc. of the Fifth Canadian Conference on General
  Relativity and Relativistic Astrophysics}, Waterloo, Ontario, May 1993
\item{[10]} Moncrief, V.: Reduction of the Einstein equations in 2+1
  dimensions to a Hamiltonian system over Teichm\"uller space, {\it
  J. Math. Phys.} 30 (1989) 2907-2914;  Mess, G.: Lorentz spacetimes of
  constant curvature, preprint Institut des Hautes \'Etudes Scientifiques
  IHES/M/90/28
 \item{[11]} Loll, R.: Independent SU(2)-loop variables and the reduced
  configuration space of SU(2)-lattice gauge theory, {\it Nucl. Phys.}
  B368 (1992) 121-142; Yang-Mills theory without Mandelstam constraints,
  {\it Nucl. Phys.} B400 (1993) 126-144
\item{[12]} Giles, R.: Reconstruction of gauge potentials from Wilson loops,
  {\it Phys. Rev.} D24 (1981) 2160-2168
\item{[13]} Loll, R.: Loop variable inequalities in gravity and gauge theory,
  {\it Class. Quant. Grav.} 10 (1993) 1471-1476
\item{[14]} Abikoff, W.: The real analytic theory of Teichm\"uller
  space,  Lecture Notes in Mathematics, vol.820, Springer, Berlin, 1980
\item{[15]} Ashtekar, A. and Romano, J.: Chern-Simons and Palatini actions
 and 2+1 gravity, {\it Phys. Lett.} B229 (1989) 56-60
\item{[16]} Gradsteyn, I.S. and Ryzhik, I.M.: {\it Table of integrals, series,
  and products}, corrected and enlarged edition, Academic Press, San Diego,
  1980
\item{[17]} Moncrief, V.: Recent advances in ADM reduction, in: {\it
  Directions in General Relativity}, vol.1, ed. B.L. Hu et al., Cambridge
  University Press, 1993, 231-243
\item{[18]} Wolpert, S.: On the Weil-Petersson geometry of the moduli
  space of curves, Ann. of Math. 115 (1982) 501-528
\item{[19]} Okai, T.: An explicit description of the Teichm\"uller space
  as  holonomy representations and its applications, Hiroshima Math. J. 22
  (1992) 259-271
\item{[20]} D'Hoker, E. and Phong, D.H.: The geometry of string perturbation
  theory, {\it Rev. Mod. Phys.} 60 (1988) 917-1065

\end